\documentclass[preprint,showkeys,secnumarabic,amsfonts,showpacs,amsmath,amssymb]{revtex4}
\usepackage[dvips]{color}
\usepackage{epsfig}
\usepackage{amsmath}
\usepackage{graphicx}
\def\Box{\hbox{$\rlap{$\sqcup$}\sqcap$}}

\begin{document}
\title{ Stability analysis in Modified Non-Local Gravity}

\author{Hossein Farajollahi}
\email{hosseinf@guilan.ac.ir} \affiliation{Department of Physics,
University of Guilan, Rasht, Iran}
\author{Farzad Milani}
\email{fmilani@guilan.ac.ir} \affiliation{Department of Physics,
University of Guilan, Rasht, Iran}
\date{\today}

\begin{abstract}
\noindent \hspace{0.35cm} In this paper we consider FRW cosmology in modified non-local gravity.
 The stability analysis shows that there is only one stable critical
point for the model and the universe undergoes a quintessence dominated era.

\end{abstract}



\maketitle

\section{Introduction}

Cosmological observations such
as Super-Nova Ia (SNIa), Wilkinson Microwave
Anisotropy Probe (WMAP), Sloan Digital Sky Survey
(SDSS), Chandra X-ray Observatory disclose some cross-checked information
of our universe \cite{Riess-A.J.(2004)}--\cite{Allen-M.N.R.A.J.(2004)}. Based on them, the universe is spatially flat,
and consists of approximately $70\%$ dark energy (DE) with negative
pressure, $30\%$ dust matter (cold dark matters plus baryons), and
negligible radiation. The combined analysis of
 SNIa, that is based upon the
background expansion history of the universe around the redshift $z
< \mathcal{O}(1)$, galaxy clusters measurements and WMAP data,
offers an evidence for the accelerated cosmic expansion
\cite{Perlmutter-A.J.(1999)}\cite{Tonry.A.J.(2003)}. The cosmological acceleration strongly
indicates that the present day universe is dominated by smoothly
distributed slowly varying DE component {\cite{Astier-A.A.(2006)}}. The
$\Lambda\text{CDM}$ model with an equation of state (EoS) parameter being $-1$ has been
continuously favored from observations. From Cosmic
Microwave Background (CMB) and Baryon Acoustic
Oscillations (BAO) \cite{Spergel-A.A.S.(2007)}\cite{Eisenstein-A.J.(2005)}, a new constraint on the EoS
parameter are observed to be around the cosmological constant value $ -1 \pm 0.1$
{\cite{Tonry.A.J.(2003)}}-{\cite{Seljak-P.R.D.(2005)}} and a possibility that
the parameter is time dependent.

The DE models with the possible phantom crossing can be broadly classified into two categories,
 (I) by adding one or more scalar field into the standard formalism of gravity, for example see
({\cite{Caldwell-P.L.B.(2002)}}-{\cite{Chiba-P.R.D.(2000)}} and refs. therein) (II) by any kind of
 modification in the geometry of the gravity, see {\cite{Capozziello-I.J.M.P.D.(2003)}}-{\cite{ Guo-P.L.B.(2005)}}.
 However, a major issue in analyzing cosmological models stems from the fact that the field equations
  are nonlinear and thus limits the possibility of obtaining exact solutions and analyzing the behavior
  of the cosmological parameters. On
the other hand, in recent years, an increasing realization of the importance of the asymptotic behavior
of cosmological models in comparison with the observational data, emphasis study of the qualitative
properties of the equations and of the long-term behavior of their solutions \cite{Setare}. In
this context, similar to the work by the authors in \cite{Jhingan} with different approach, we use perturbation method, especially linear stability and phase-plane
analysis to study the stability of a modified non local cosmological model and investigate the possible quintessence scenarios in the formalism.

The paper is organized as follows: Section two is concerned with the dynamics of the FRW cosmology in modified non local gravity. In section three we study the stability of the autonomous system of
differential equations.
We consider a linear perturbation for our model and investigate the evolution
of the cosmological perturbations and identify the
attractor solution for the perturbations during
the tracking regime. We also examine the dynamics of the universe by stability analysis. Finally, we summaries our paper in section four .

\section{The Model}

We start with the action of the non-local gravity as a simple
modified garvity given by {\cite{Nojiri-P.L.B.(2008)}},
\begin{eqnarray}\label{ac1}
S=\int{d^4x\sqrt{-g}\left\{\frac{M^{2}_{p}}{2}R(1+f(\Box^{-1}R))\right\}}\cdot
\end{eqnarray}
where $M_{p}$ is Plank mass, $f$ is some function and $\Box$ is
d'Almbertian for scalar field. Generally speaking, such non-local
effective action, derived from string theory, may be induced by quantum effects. A Bi-
scalar reformation of non-local action can be presented by introducing two scalar
fields $\phi$ and $\psi$, where changes the above action to a local
from:
\begin{eqnarray}
S&=&\int{d^{4}x\sqrt{-g}\left\{\frac{M^{2}_{p}}{2}\left[R(1+f(\phi))+\psi(\Box\phi-R)\right]\right\}},\label{ac2}
\label{ac}
\end{eqnarray}
where $\psi$, at this stage, plays role of a lagrange multiplier. One might
further  rewrite the above action as
\begin{eqnarray}
S&=&\int{d^{4}x\sqrt{-g}\left\{\frac{M^{2}_{p}}{2}\left[R(1+f(\phi)-\psi)-\partial_
{\mu}\psi\partial^{\mu}\phi\right]\right\}},\cdot\label{ac}
\end{eqnarray}
which now is equivalent to a local model with two extra degrees of
freedom. By the variation over $\psi$, we obtain $\Box\phi=R$ or
$\phi= \Box^{-1}R$.

  Now in a FRW cosmological model with
 invariance of the action under changing fields and
vanishing variations at the boundary, the equations of motion for
only time dependent scalar fields, $\phi$ and $\psi$,  become
\begin{eqnarray}
\ddot{\phi}+3H\dot{\phi}+R&=&0,\label{EoS1}\\
\ddot{\psi}+3H\dot{\psi}-Rf'&=&0,\label{EoS2}
\end{eqnarray}
where the scalar curvature $R$ is $R=12H^2+6\dot{H}$, $H$ is Hubble parameter and prime means derivative with respect to the scalar field.
Variation of action (\ref{ac}) with respect to the metric tensor $g_{\mu\nu}$ gives,
\begin{eqnarray}\label{T}
0&=&\frac{1}{2}g_{\mu\nu}\left\{R(1+f-\psi)-\partial_{\rho}\psi\partial^{\rho}\phi\right\}-R_{\mu\nu}(1+f-\psi)
+\frac{1}{2}(\partial_{\mu}\psi\partial_{\nu}\phi+\partial_{\mu}\phi\partial_{\nu}\psi)\nonumber\\
&-&(g_{\mu\nu}\Box-\nabla_{\mu}\nabla_{\nu})(f-\psi)\cdot
\end{eqnarray}
The $00$ and $ii$ components of the equation (\ref{T}) are
 \begin{eqnarray}
0&=&-3H^2(1+f-\psi)+\frac{1}{2}\dot{\psi}\dot{\phi}-3H(f'\dot{\phi}-\dot{\psi}),\label{f1}\\
0&=&(2\dot{H}+3H^2)(1+f-\psi)+\frac{1}{2}\dot{\psi}\dot{\phi}+(\frac{d^2}{dt^2}+2H\frac{d}{dt})(f-\psi)\cdot\label{f2}
\end{eqnarray}
Equations (\ref{f1}) and (\ref{f2}) can be rewritten as
\begin{eqnarray}
3H^2&=&\frac{\frac{1}{2}\dot{\psi}\dot{\phi}-3H(f'\dot{\phi}-\dot{\psi})}{(1+f-\psi)},\label{f3}\\
2\dot{H}+3H^2&=&-\frac{\frac{1}{2}\dot{\psi}\dot{\phi}+(\frac{d^2}{dt^2}+2H\frac{d}{dt})(f-\psi)}{(1+f-\psi)}\cdot\label{f4}
\end{eqnarray}
Comparison with the standard Friedman equations, the right hand side of the equations (\ref{f3}) and (\ref{f4}) can be treated as the effective energy density and pressure:
\begin{eqnarray}
\frac{\rho_{eff}}{M_{p}^{2}}&=&\frac{\frac{1}{2}\dot{\psi}\dot{\phi}-
3H(f'\dot{\phi}-\dot{\psi})}{(1+f-\psi)},\label{f5}\\
\frac{p_{eff}}{M_{p}^{2}}&=&\frac{\frac{1}{2}\dot{\psi}\dot{\phi}+(\frac{d^2}{dt^2}+2H\frac{d}{dt})
(f-\psi)}{(1+f-\psi)}\cdot\label{f6}
\end{eqnarray}
Using Eqs. (\ref{EoS1}) and (\ref{EoS2}) and doing some algebraic calculation we can read the effective
 energy density and pressure from the above as,
\begin{eqnarray}\label{rho}
\rho_{eff}=\frac{M^{2}_{p}}{1+f-\psi}\left\{\frac{1}{2}\dot{\psi}\dot{\phi}-3H(f'\dot{\phi}-\dot{\psi})\right\}\cdot
\end{eqnarray}
\begin{eqnarray}\label{p}
p_{eff}=\frac{M^{2}_{p}}{1+f-\psi-6f'}\left\{\frac{1}{2}
\dot{\psi}\dot{\phi}+f''\dot{\phi}^{2}-
H(f'\dot{\phi}-\dot{\psi})+\frac{f'\left[6H(f'\dot{\phi}-\dot{\psi})
-\dot{\psi}\dot{\phi}\right]}{1+f-\psi}\right\}\cdot
\end{eqnarray}
Now by using Eqs. (\ref{rho}) and (\ref{p}) the conservation
equation can be obtained as,
\begin{eqnarray}\label{rhodot}
\dot{\rho}_{eff}+3H\rho_{eff}(1+\omega_{eff})=0,
\end{eqnarray}
where $\omega_{eff}$ is the EoS parameter of the model.

\section{perturbation and Stability}
In this section, we study the structure of the dynamical system via  phase plane analysis,
by introducing the following four dimensionless variables,
\begin{eqnarray}
x=-f(\phi),\ \  y=\frac{\dot{\phi}}{H},\ \  z=\frac{\dot{\psi}}{6H},\ \  k=\psi.
\end{eqnarray}
From Friedmann Eq.(\ref{f3}) and assuming $f(\phi)=f_0 e^{\sigma\phi}$, one finds the Friedmann constraint equation in terms of the new dynamical variables as
\begin{eqnarray}\label{per1}
1=x+k+zy+\sigma xy+6z.
\end{eqnarray}
Then using equations (\ref{EoS1})--(\ref{EoS2}), (\ref{f3})-(\ref{f4}), and (\ref{rhodot}) the evolution equations of these variables become,
\begin{eqnarray}
x'&=&\frac{dx}{d\ln a}=\frac{\dot{x}}{H}=\sigma yx,\label{xprime}\\
y'&=&\frac{dy}{d\ln a}=\frac{\dot{y}}{H}=-(3+\frac{\dot{H}}{H^2})y-6(2+\frac{\dot{H}}{H^2}),\label{yprime}\\
z'&=&\frac{dz}{d\ln a}=\frac{\dot{z}}{H}= -(3+\frac{\dot{H}}{H^2})z-\sigma(2 x + \frac{\dot{H}}{H^2}),\label{zprime}\\
k'&=&\frac{dk}{d\ln a}=\frac{\dot{k}}{H}= 6z.\label{kprime}
\end{eqnarray}
Also, we can obtain effective EoS parameter and deceleration parameter as $q=-1-\frac{\dot{H}}{H^2}$, $\omega_{eff}=-1-\frac{2}{3}\frac{\dot{H}}{H^2}$,
where in terms of the new dynamical variables we have,
\begin{eqnarray}\label{HdotH2}
\frac{\dot{H}}{H^2}=-\frac{4\sigma x(6+y)+4z-\sigma^2 xy^2+6zy}{2\left(1-x(1-6\sigma)-k\right)}\cdot
\end{eqnarray}
For an expanding universe with a scale factor $a(t)$ given by $a \propto t^p$ one also can find
\begin{eqnarray}\label{pstability}
p=\frac{2\left(1-x(1-6\sigma)-k\right)}{4\sigma x(6+y)+4z-\sigma^2 xy^2+6zy},
\end{eqnarray}
in terms of the new dynamical variables. We will restrict our discussion on the existence and stability of critical points in an expanding
universes with $H>0$. Critical (fixed) points correspond to points where $x'=0$, $y'=0$, $z'=0$ and $k'=0$.

In order to study the stability of the critical points, using the Friedman constraint
equation (\ref{per1}) we first reduce Eqs.(\ref{xprime})-(\ref{kprime}) to three independent equations for $x$, $y$ and $z$. Substituting
linear perturbations $x\rightarrow x+\delta x$, $y\rightarrow y+\delta y$ and $z\rightarrow z+\delta z$ about
the critical points into the above three independent equations to first-order in the perturbations,
gives the evolution equations of the linear perturbations, which yield three eigenvalues $\lambda_i$.
Stability requires the real part of all eigenvalues to be negative.  The linearization of the system about these fixed points yields three eigenvalues
\begin{eqnarray}
\lambda_1=-\frac{25}{9},\,\,\,\,\,\,\ \lambda_2=-\frac{4}{9},\,\,\,\,\,\,\ \lambda_3=-\frac{4}{9}. \nonumber
\end{eqnarray}
After solving the new set of equations we find that there is only one critical point for our system with the physical property presented in table 1. As stated the critical point is a stable point in our model for $\sigma=-\frac{20}{729}$.

\begin{table}[ht]
 % title of Table
\centering % used for centering table
\begin{tabular}{c c c c c c c c c c} % centered columns (10 columns)
\hline\hline %inserts double horizontal lines
$\textbf{Label}$\,\,\,\,\,\,\,\,\, & $\textit{\textbf{x}}$\,\,\,\,\,\,\,\,\, & $\textit{\textbf{y}}$\,\,\,\,\,\,\,\,\,
 & $\textit{\textbf{z}}$\,\,\,\,\,\,\,\,\, &  $\omega_{eff}$\,\,\,\,\,\,\,\,\,& $\textit{\textbf{q}}$\,\,\,\,\,\,\,\,\,
 &$\textit{\textbf{p}}$\,\,\,\,\,\,\,\,\,&\textbf{Stability}\\ [0.5 ex] % inserts table
%heading
\hline\hline % inserts single horizontal line
\textbf{S}\,\,\,\,\,\,\,\,\, & 1 \,\,\,\,\,\,\,& 0 \,\,\,\,\,\,\,& 0\,\,\,\,\,\,\,\,\,&  $\frac{1}{3}$ \,\,\,\,\,\,\,\,\,
 & 1 \,\,\,\,\,\,\,\,\,& $\frac{1}{2}$ \,\,\,\,\,\,\,\,\,& stable \\ % inserting body of the table
\hline %inserts single line
\end{tabular}
\caption{The properties of the critical point.}
\label{table:nonlin} % is used to refer this table in the text
\end{table}

The stable critical point, S, represents a scalar field-dominated solution in the radiation dominated evolutive universe with $a \propto t^{\frac{1}{2}}$. In Fig. 1, the attraction of trajectories to the critical point in the three dimensional phase plane is shown for the given initial
conditions.

\begin{tabular*}{2.5 cm}{cc}
\includegraphics[scale=.5]{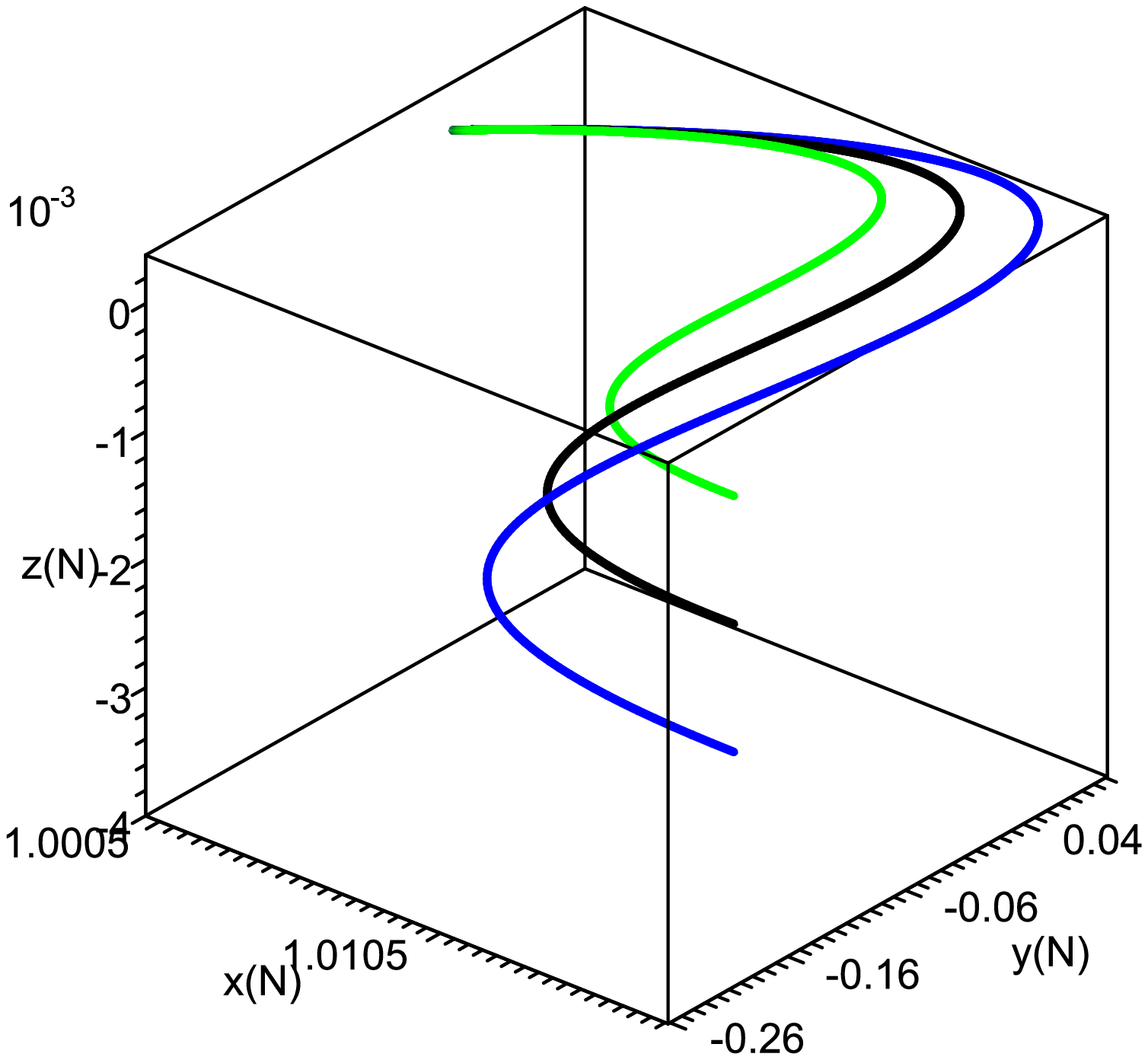}\\
Fig. 1: \, The attractor property of the dynamical system in
the phase plane.\\ Initial values are:
(blue): $x(0)=1.0099$, \,$y(0)=0.002$, \,$z(0)=-0.004$.\\
(black): $x(0)=1.0099$, \,$y(0)=0.002$, \,$z(0)=-0.003$.\\
(green): $x(0)=1.0099$, \,$y(0)=0.002$, \,$z(0)=-0.002$.\\
\end{tabular*}\\

Fig. 2 shows the projection of the three dimensional phase space onto the two dimensional $y=0.002$ phase space in terms of the dynamical variables $x$ and  $z$.\\

\begin{tabular*}{2.5 cm}{cc}
\includegraphics[scale=.35]{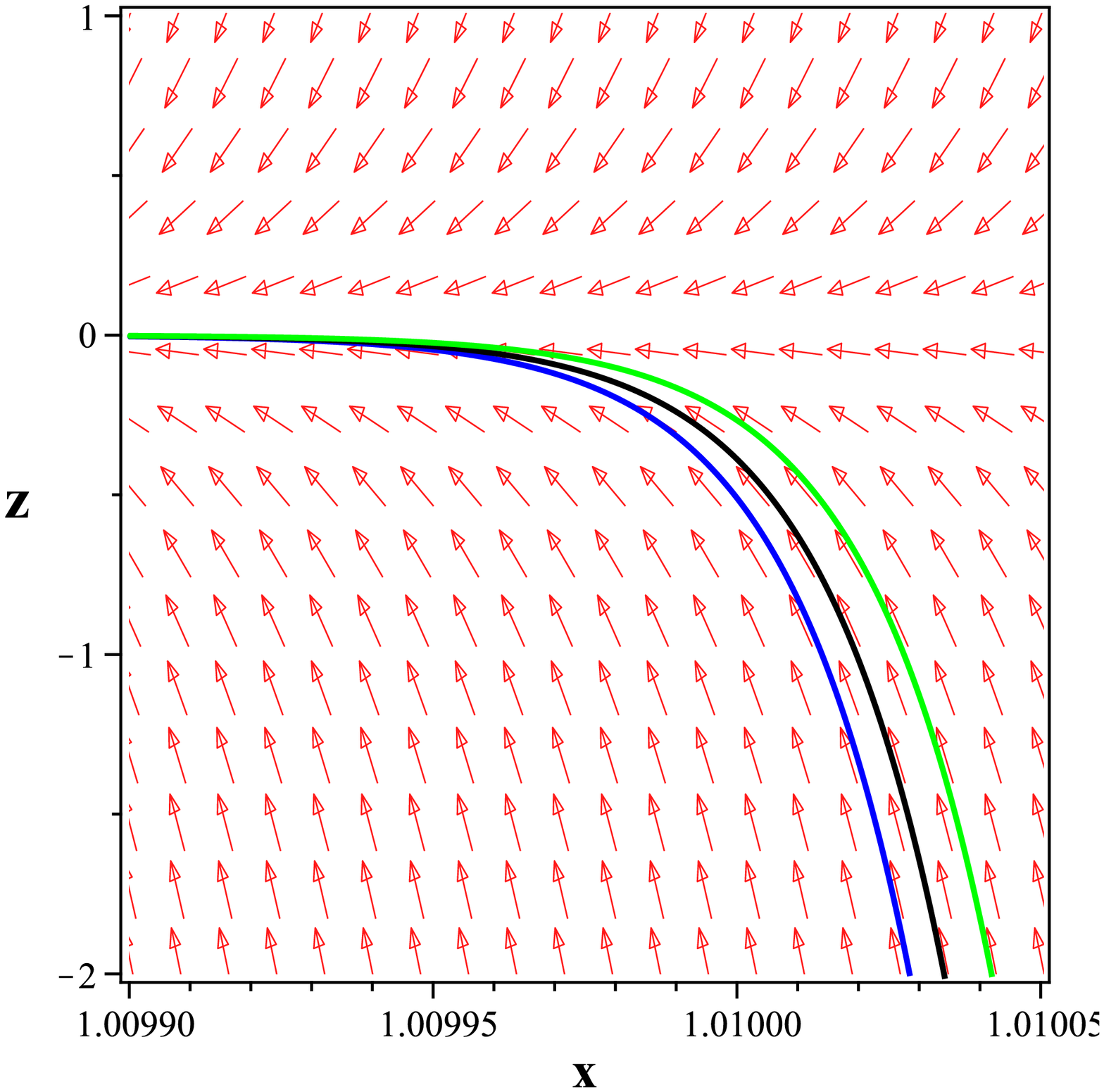}\hspace{0.2 cm}\\
\hspace{0.5 cm}Fig. 2: \, The phase plane of variation of $z$ respect of $x$, asymptotically $y=0.002$ \\
stable equilibrium sink. Initial values are: (blue): $x(0)=1.0099$, \, $z(0)=-0.004$.\\
(black): $x(0)=1.0099$, \,$z(0)=-0.003$.
(green): $x(0)=1.0099$, \,$z(0)=-0.002$.\\
\end{tabular*}\\

Fig. 3 shows the projection of the three dimensional phase space onto the two dimensional $x=1.01$ phase space in terms of the dynamical variables $y$ and  $z$.\\

\begin{tabular*}{2.5 cm}{cc}
\includegraphics[scale=.35]{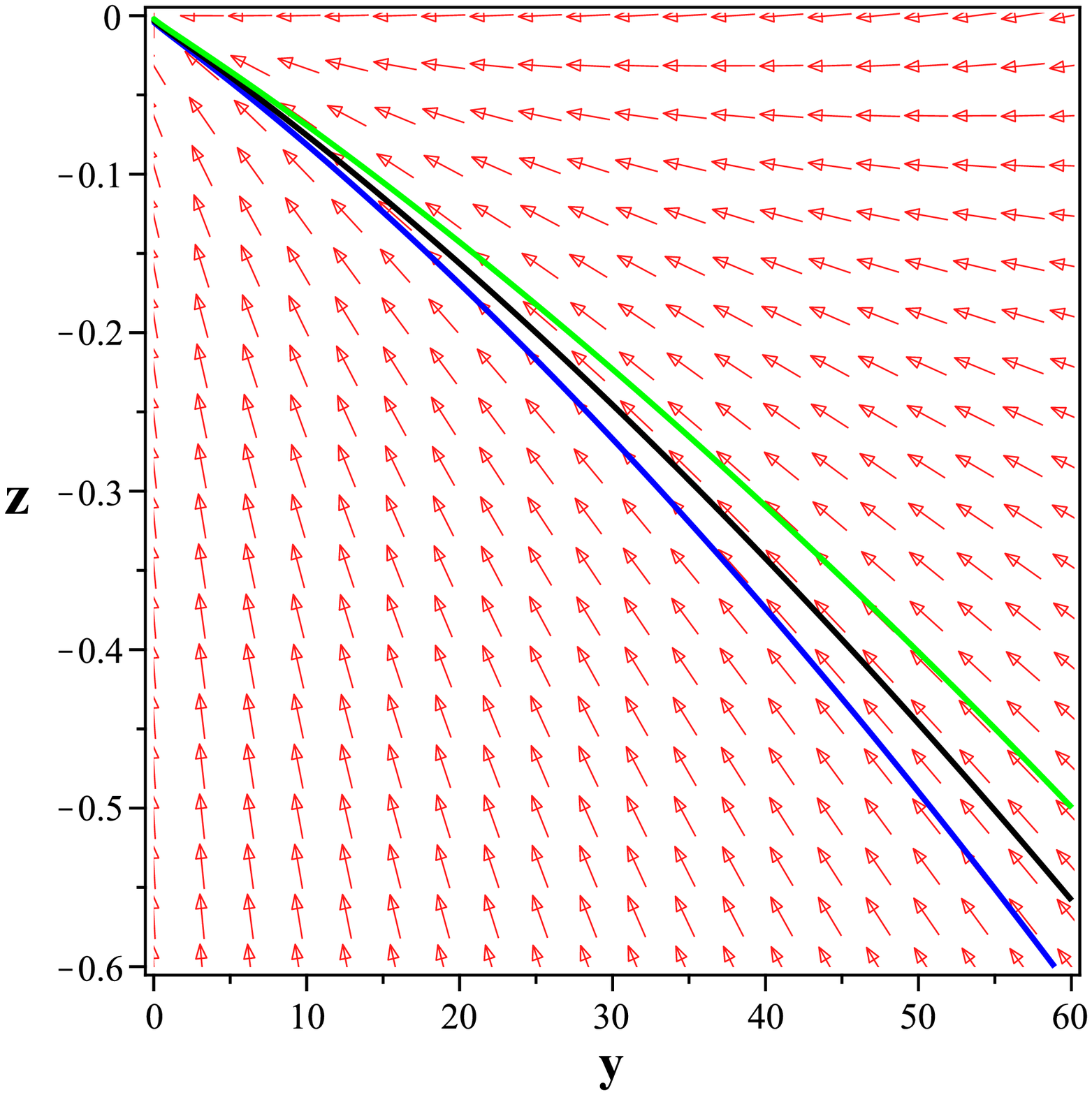}\\
\hspace{0.5 cm}Fig. 3: \, The phase plane of variation of $z$ respect of $y$, asymptotically $x=1.01$ \\
stable equilibrium sink. Initial values are: (blue): $y(0)=0.002$, \, $z(0)=-0.004$.\\
(black): $y(0)=0.002$, \,$z(0)=-0.003$.
(green): $y(0)=0.002$, \,$z(0)=-0.002$.\\
\end{tabular*}\\

As can be seen, in Figs. 2 and 3 all the trajectories for different values of initial conditions asymptotically approaches the projected stable critical point in their plane. It means that in a dynamical system sense, this critical point is an attractor and the system is drawn to this point in a self-organizing manner.

At this stage we study the cosmological evolution of EoS parameter, $\omega_{eff}$ in terms of the new dynamical variables. As can be seen in Fig. 4, for the stable critical point, by perturbation, the trajectory of the EoS parameter with the given initial conditions near the critical point values, starts from the state in the past with $\omega_{eff}=1/3$, undergoes a quintessence dominated era with $-1<\omega_{eff}<0$ sometimes in that past and future, passes the current negative value $\omega_{eff}\simeq -0.14$ and finally approaches the same state with $\omega_{eff}=1/3$ in the future, which correspond to the stable critical point (blue curve). In this Figure we also show, for completeness, that if the the initial conditions are far enough from the critical point values, by perturbation, the universe starts from the state in the past with $\omega_{eff}=1/3$ and undergoes phantom crossing twice in the past and future and never come pack to the same state in the future ( the red curve).

\begin{tabular*}{2.5 cm}{cc}
\includegraphics[scale=.4]{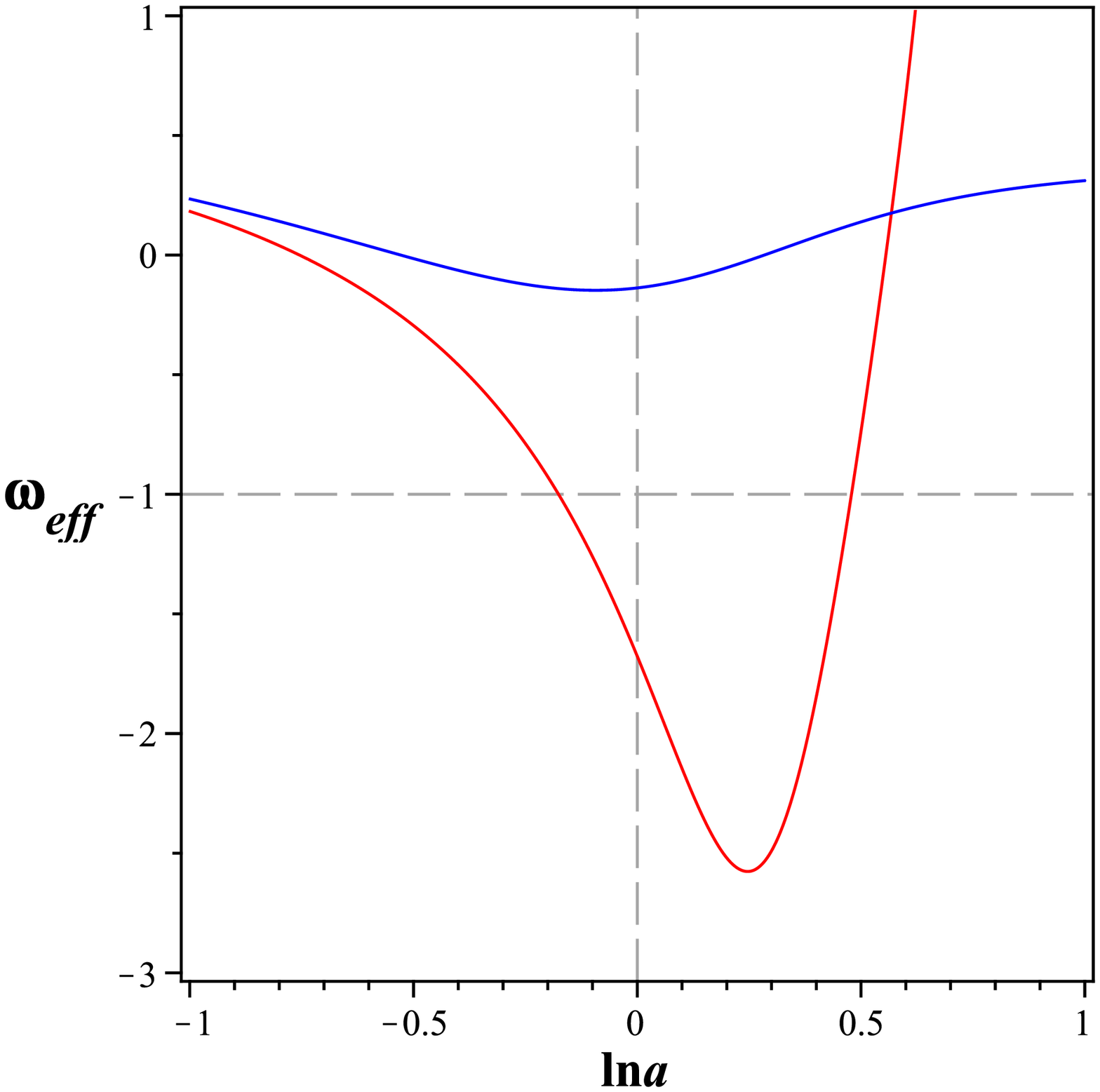}\\
\hspace{2.0 cm}Fig. 4: \, The graph of $\omega_{eff}$ respect of $\ln(a)$. Initial values are: \\
\hspace{2.0 cm}(red): $x(0)=-2$, \,$y(0)=-2.5$,  \, $z(0)=0.3$,\\
\hspace{2.0 cm}(blue): $x(0)=1.01$,\, $y(0)=-0.04$, \,$z(0)=-0.02$,\\
\end{tabular*}\\

\section{Summary and Remarks}

In this paper, we consider a local scalar-tensor formulation of non-local gravity as a simple modified model
 characterized by two scalar fields $\phi$ and $\psi$
and function $f(\phi)$ which can be viewed as scalar potential in the model.

We have made a perturbative analysis of the cosmological model,
to investigate the stability of the model. The perturbative equations of motion are solved numerically,
and we found that the system is stable for only one critical point under scalar perturbation.
The three dimensional phase space of the model gives the corresponding conditions
for tracking attractor. We have also shown the projected two dimensional phase space in $x=constant$ and $y=constant$ planes.
The stability analysis shows that, for the only stable critical point, the phantom crossing never occurs. After perturbation, the universe approaches a stable attractor in the future with $\omega_{eff}=1/3$ (radiation dominated era) and $p=1/2$ (decelerating expansion era). The universe may undergoes phantom crossing if the chosen initial condition are far enough from the critical point values, such that the universe never come back to the initially stable state.

\end{document}